\documentclass[prl,aps,twocolumn,showpacs]{revtex4}
\usepackage{graphicx}
\usepackage{amsmath}
\usepackage{bm}
\usepackage{amstext}
\usepackage{amsxtra}

\usepackage{times}

\begin{document}

\newcommand{\To}{T_c^0}
\newcommand{\kB}{k_{\rm B}}
\newcommand{\dT}{\Delta T_c}
\newcommand{\lo}{\lambda_0}
\newcommand{\cs}{$\clubsuit$}
\newcommand{\thold}{t_{\rm hold}}
\newcommand{\Nmf}{N_c^{\rm MF}}
\newcommand{\Tmf}{T_c^{\rm MF}}
\newcommand{\bra}[1]{\langle #1|}
\newcommand{\ket}[1]{|#1\rangle}
\newcommand{\downstate}{\left\vert\downarrow\right\rangle}
\newcommand{\upstate}{\left\vert\uparrow\right\rangle}
\newcommand{\Ndown}{N_{\downarrow}}
\newcommand{\Nup}{N_{\uparrow}}
\newcommand{\tc}{t_{\rm coh}}

\title{
Persistent currents in spinor condensates
}

\author{Scott Beattie, Stuart Moulder, Richard J. Fletcher, and Zoran Hadzibabic}
\affiliation{Cavendish Laboratory, University of Cambridge, J.~J.~Thomson Ave., Cambridge CB3~0HE, United Kingdom}

\begin{abstract}
We create and study persistent currents in a toroidal two-component Bose gas, consisting of $^{87}$Rb atoms in two different spin states. For a large spin-population imbalance we observe supercurrents persisting for over two minutes.
However we find that the supercurrent is unstable for spin polarisation below a well defined critical value. We also investigate the role of phase coherence between the two spin components and show that only the magnitude of the spin-polarisation vector, rather than its orientation in spin space, is relevant for supercurrent stability.
\end{abstract}

\date{\today}

\pacs{03.75.Kk, 67.85.-d}


\maketitle



Persistent currents are a hallmark of superfluidity and superconductivity, and have been studied in liquid helium and solid state systems for decades. More recently, it became possible to trap an atomic Bose-Einstein condensate (BEC) in a ring geometry \cite{Gupta:2005,Arnold:2006,Ryu:2007,Henderson:2009,Ramanathan:2011,Sherlock:2011, Moulder:2012, Wright:2012} and induce rotational superflow in this system \cite{Ryu:2007,Ramanathan:2011, Moulder:2012, Wright:2012}. This offers new possibilities for fundamental studies of superfluidity in a flexible experimental setting. Both long-lived superflow \cite{Ramanathan:2011,Moulder:2012} and quantised phase slips corresponding to singly-charged vortices crossing the superfluid annulus have been observed \cite{Moulder:2012,Wright:2012}. 

So far experiments on persistent currents in atomic BECs were limited to spinless, single-component condensates. Extending such studies to multi-component systems, in particular those involving two or more spin states \cite{Ho:1998, Ohmi:1998, Stenger:1998b}, is essential for understanding superfluids with a vectorial order parameter and for applications in atom interferometry \cite{Gustavson:1997,Halkyard:2010}.
Persistent flow in a two-component Bose gas has been studied theoretically \cite{Ho:1982, Smyrnakis:2009, Bargi:2010,Anoshkin:2012} but many issues remain open. Even the central question of whether, and under what conditions, this system supports persistent currents has not been settled.

In this Letter, we study the stability of supercurrents in a toroidal  two-component gas consisting of $^{87}$Rb atoms in two different spin states. For a large spin-population imbalance we observe superflow persisting for over two minutes and limited only by the atom-number decay. However at a small population imbalance the onset of supercurrent decay occurs within a few seconds. We demonstrate the existence of a well defined critical spin-polarisation separating the stable- and unstable-current regimes. We also study the connection between spin coherence and superflow stability, and show that in our system only the modulus of the spin-polarisation vector is relevant for the stability of the supercurrent.
The existence of a critical population imbalance was anticipated in Refs.~\cite{Smyrnakis:2009, Bargi:2010,Anoshkin:2012}, but quantitative comparison with our measurements will require further theoretical work.

\begin{figure}[bp]
\includegraphics[width=\columnwidth]{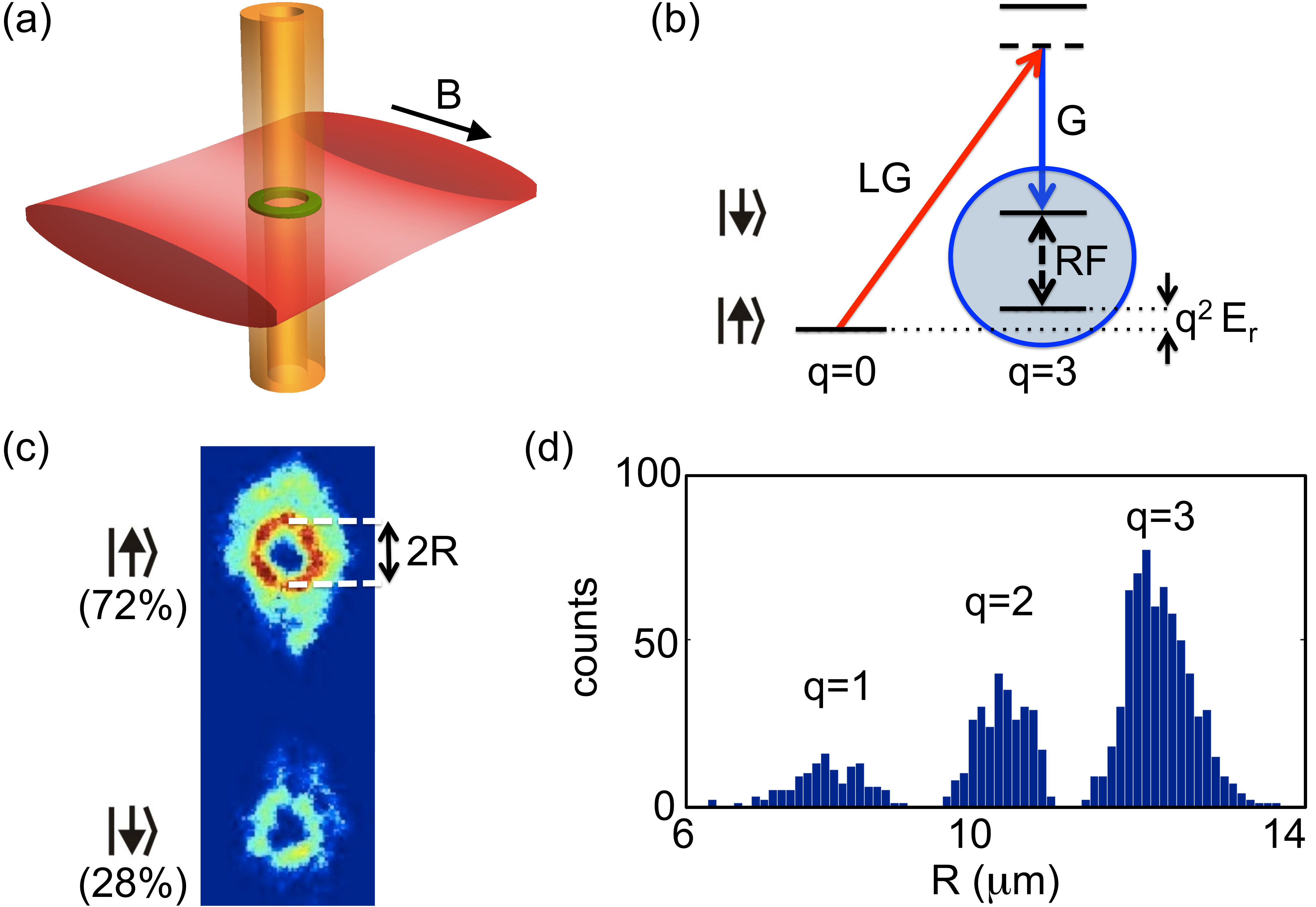}
\caption{(color online) Preparation and detection of supercurrent in a two-component gas.
(a) The ring trap is formed by a horizontal ``sheet" beam and a vertical Laguerre-Gauss (LG) ``tube" beam. $B$ is the external magnetic field.
(b) Supercurrent is induced by a Raman transfer of atoms between two spin states, $\upstate$ and $\downstate$, using the LG beam and an auxiliary Gaussian (G) beam. 
During the transfer each atom absorbs $3\hbar$ of angular momentum from the LG beam. 
Two-component gas is created by coupling $\upstate$ and $\downstate$ states with an RF field. 
The characteristic rotational energy is $E_r /h \approx 0.4\;$Hz.
(c) Time-of-flight image of the atoms, with spin states separated using a Stern-Gerlach gradient. The rotational state $q$ is deduced from the radius $R$ characterising the central hole in the density distribution. 
The image shown was taken after $t=4$~s of rotation; the longitudinal spin-polarisation is $P_z = 0.44$ and $q=3$ for both spin states.
(d) Histogram of $\approx 900$ measurements of $R$ at various $P_z$ and $t$. 
}
\label{fig:exp}
\end{figure}

Our setup is outlined in Fig.~\ref{fig:exp}(a). We load a BEC of $N \approx 10^5$  atoms into an optical ring trap of radius 12~$\mu$m, created by intersecting a 1070~nm ``sheet" laser beam and an 805~nm ``tube" beam \cite{Moulder:2012}. 
The sheet beam confines the atoms to the horizontal plane with a trapping frequency of 350~Hz. In-plane, the tube beam confines the atoms to the ring with a trapping frequency of 50~Hz. 
The trap depth is about twice the BEC chemical potential, $\mu_0/h \approx 0.6\;$kHz, and varies azimuthally by $< 10\%$.

Our tube trapping beam is a Laguerre-Gauss LG$^3$ laser mode in which each photon carries orbital angular momentum $3 \hbar$. We use the same beam to induce a supercurrent via a two-photon Raman process \cite{Andersen:2006,Ryu:2007,Moulder:2012}.  We briefly ($200\;\mu$s) pulse on an auxiliary TEM$_{00}$ Gaussian beam, copropagating with the LG beam, to transfer all atoms between two spin states, $\upstate$ and $\downstate$ [Fig.~\ref{fig:exp}(b)]. Each atom absorbs angular momentum $3 \hbar$ from the LG beam and
we thus create a (single-component) current corresponding to a vortex of charge $q=3$ trapped at the ring centre. Such current can persist for over a minute and decays in quantised $q \rightarrow q-1$ steps, corresponding to $2 \pi$ phase slips in the BEC wave function \cite{Moulder:2012}.

The $\upstate$ and $\downstate$ states also define the spin space for our two-component experiments. 
To create a two-component current we prepare a pure $\vert q=3, \downarrow \rangle$ state and then couple $\upstate$ and $\downstate$ by a radio-frequency (RF) field, which carries no orbital angular momentum and does not affect the motional state of the atoms. 
The $\upstate$ and $\downstate$ are two $F= 1$ hyperfine ground states, $m_F = 1$ and 0, respectively. The $m_F = -1$ state is detuned from Raman and RF resonances by the quadratic Zeeman shift in an external magnetic field $B$ of 10 gauss.

After preparing a rotating ($q=3$) cloud in a specific spin state, 
we let it evolve in the ring trap for a time $t$ and then probe it by absorption imaging after 29~ms of time-of-flight expansion. We separate the two spin components with a Stern-Gerlach gradient and directly measure the longitudinal spin-polarisation $P_z =  (\Nup - \Ndown)/(\Nup + \Ndown)$, where $\Nup$ ($\Ndown$) is the number of atoms in the $\upstate$ ($\downstate$) state [Fig.~\ref{fig:exp}(c)].
The rotational state, $0\leq q \leq 3$, is seen in the size $R$ of the central hole in the atomic distribution \cite{Moulder:2012}, arising due to a centrifugal barrier \cite{Ryu:2007}. As shown in Fig.~\ref{fig:exp}(d), the $R$ values are clearly quantised \cite{Moulder:2012, Wright:2012}, allowing us to determine $q$ with $>99\%$ fidelity \cite{bending}.

\begin{figure}[tbp]
\includegraphics[width=\columnwidth]{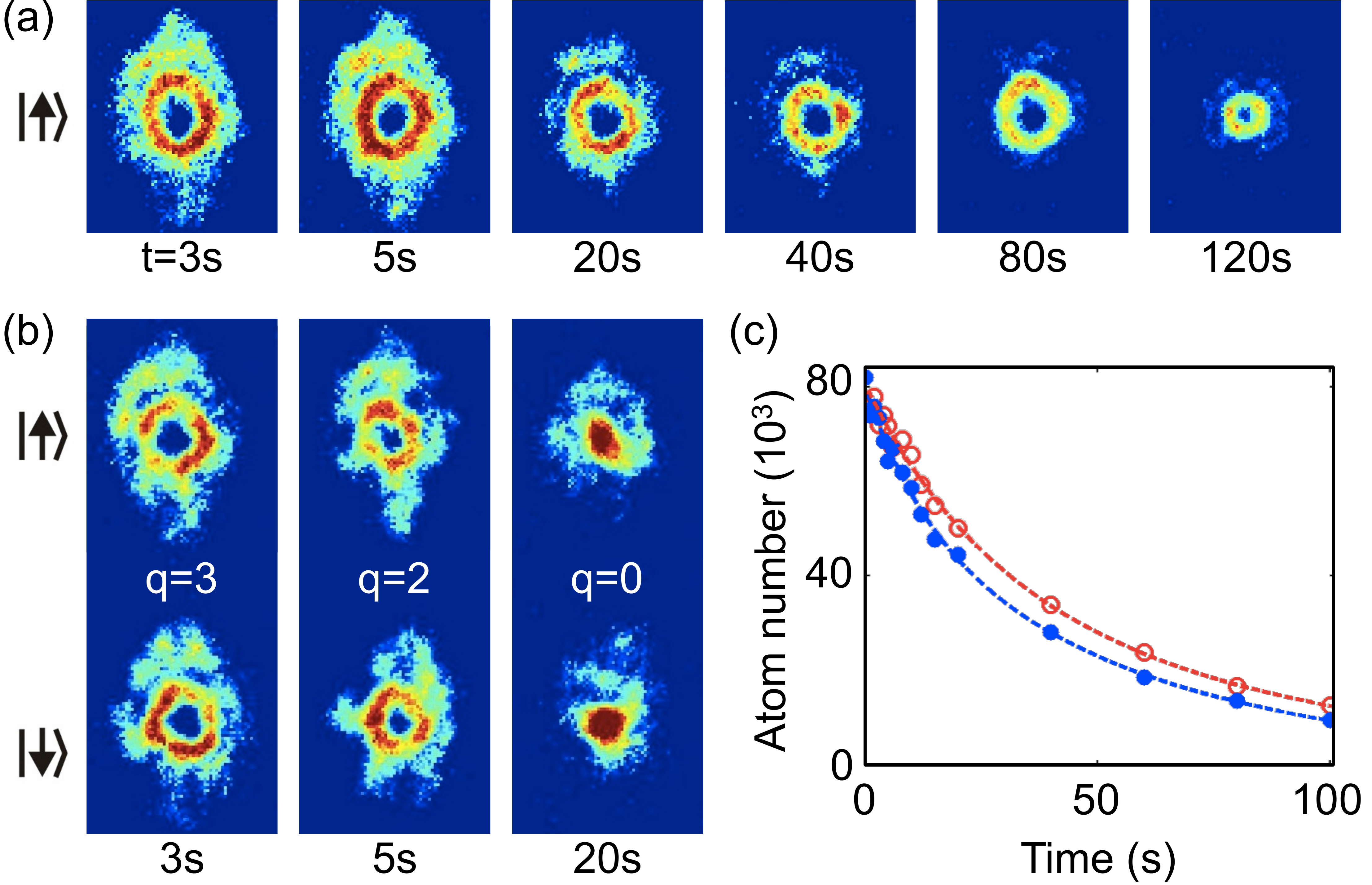}
\caption{(color online) Single- vs. two-component supercurrent. (a) In a pure $\upstate$ state ($P_z=1$) supercurrent persists for over two minutes, with no phase slips occurring for $\approx 90$~s. (b) At $P_z=0$ the first phase slip occurs within 5~s and we observe no rotation beyond 20~s. (c) Total atom number decay for $P_z=1$ (open symbols) and  $P_z=0$ (solid symbols). Dashed lines are double-exponential fits.
}
\label{fig:movie}
\end{figure}

In Fig.~\ref{fig:movie} we illustrate the dramatic difference between superflow stability in a $P_z=1$ single-component gas and a $P_z=0$ two-component system. 
The two different $P_z$ states are created, respectively, by a ($140\,\mu$s) $\pi$ and a ($70\,\mu$s) $\pi/2$ RF pulse at $t=0$.
In the pure $\upstate$ state [Fig.~\ref{fig:movie}(a)] the current persists for over two minutes, with the BEC always remaining in the $q=3$ state for $\approx 90$~s. 
In contrast, at $P_z =0$ [Fig.~\ref{fig:movie}(b)] the first phase slip occurs within 5~s and the current completely decays within 20~s. During the decay we always observe the two spin components to be in the same $q$ state. 

Supercurrent stability generally depends on the number of condensed atoms \cite{Ramanathan:2011, Moulder:2012} and at $P_z=0$ the atom number per spin state is halved. However, from the $N$-decay curves in Fig.~\ref{fig:movie}(c) we see that this alone cannot explain the difference in superflow stability. At $P_z=1$ rotation still persists for $N \approx 10^4$ while at $P_z=0$ it stops already at $N > 4 \times 10^4$.
Moreover, if we apply a $\pi/2$ RF pulse at $t=0$ but then immediately remove all the $\upstate$ atoms from the trap with a resonant light pulse, the current again persists for over a minute. This unambiguously confirms that in Fig.~\ref{fig:movie}(b) the superflow is inhibited by the presence of {\it both} spin components.

\begin{figure}[tbp]
\includegraphics[width=\columnwidth]{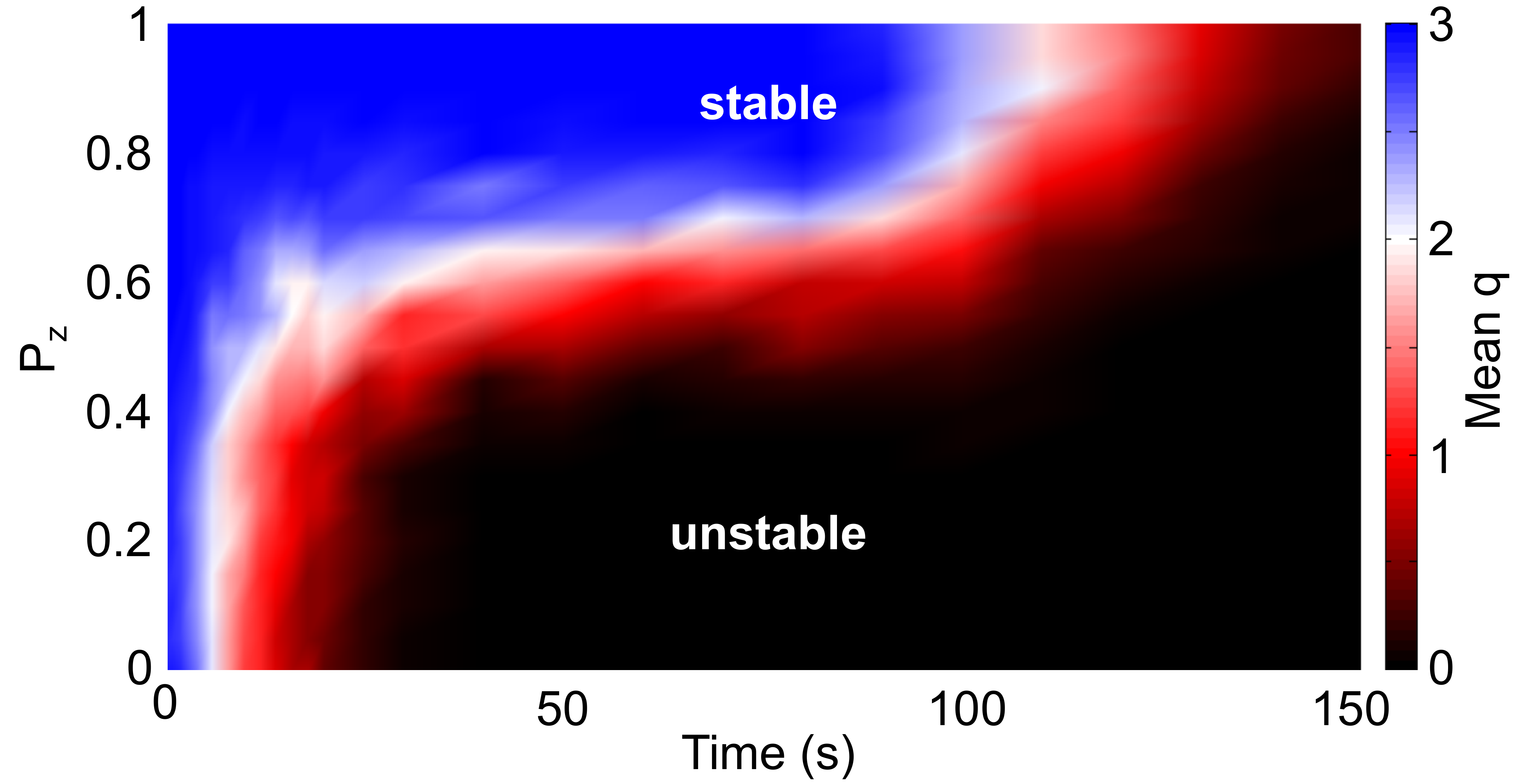}
\caption{(color online) Supercurrent stability in a partially spin-polarised gas. 
The statistically averaged supercurrent state, $\langle q \rangle$, of the majority spin component is shown as a function of $P_z$ and the evolution time $t$. The contour plot is based on $\approx 1600$ measurements of $q(P_z,t)$. The transition between stable- and unstable-current regimes occurs at $0.6<P_z<0.7$. In the stable regime the current eventually decays due to the atom-number decay.
}
\label{fig:surface}
\end{figure}

We now turn to a quantitative study of the supercurrent stability as a function of the spin-population imbalance (Fig.~\ref{fig:surface}). 
We tune $P_z$ by varying the length $\Delta t$ of the RF pulse applied at $t=0$, and measure the $q$ state of the majority $\left( \upstate \right)$ spin component as a function of $t$. Whenever the radius $R$ is fittable for the minority component we get the same $q$ for both spin components in $>99\,\%$ of cases. However for $\Ndown <10^4$ we cannot determine $q$ for the minority component.

Based on $\approx 1600$ measurements of $q(P_z, t)$, in Fig.~\ref{fig:surface} we reconstruct the complete current stability diagram for $0 \leq P_z \leq 1$ \cite{symmetry}. The contour plot of $\langle q (P_z, t) \rangle$ is obtained by spline interpolation through a 3D mesh of data points with integer $q$ values.
The blue-shaded region corresponds to rotation times for which no phase slips occur.  We clearly distinguish two qualitatively different regimes. For large $P_z$ the superflow is fundamentally stable and limited only by the atom-number decay; for low $P_z$ the current starts to decay within a few seconds. A sharp transition between the two regimes occurs at $0.6 < P_z <0.7$.

\begin{figure}[tbp]
\includegraphics[width=\columnwidth]{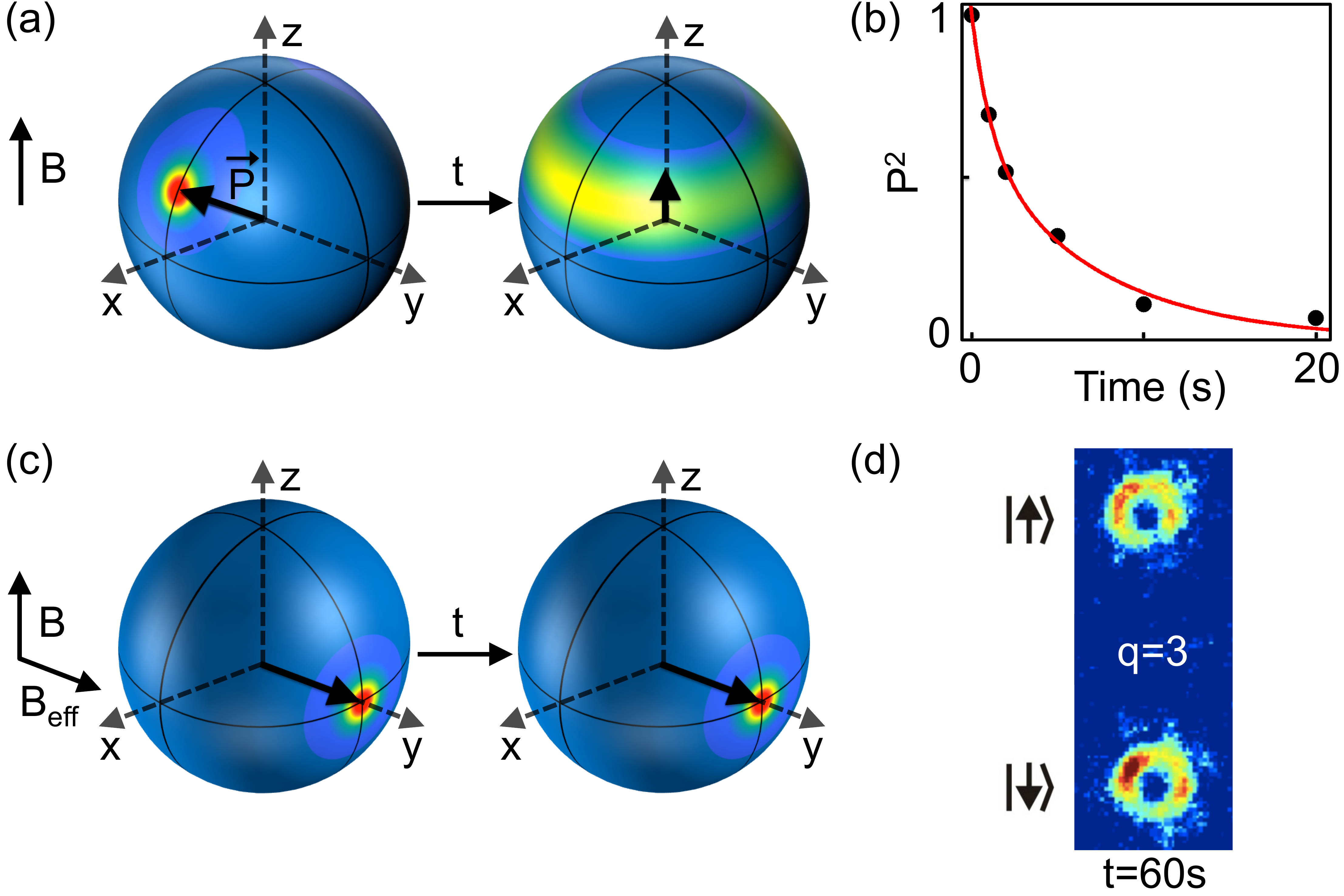}
\caption{(color online) Role of spin-coherence in superflow stability.
(a) An RF pulse at $t=0$ creates a spin-superposition state in which $P_z < 1$ but $P=1$. As the superposition decoheres, transverse spin-polarisation decays and $P \rightarrow P_z$. 
(b) Transverse-polarisation decay (see text).  Double-exponential fit (solid red line) gives the decay function $f(t)$.
(c) Adiabatic dressing of the spin state. In presence of a resonant RF field $\vec{B}_{\rm eff} \propto \hat{y}$ and the dressed $P_z=0$ state $|y\rangle$ is stable against decoherence.
(d) In the dressed $|y\rangle$ state the supercurrent is also stable.}
\label{fig:spin}
\end{figure}

To fully understand these observations, we need to distinguish a coherent superposition of $\upstate$ and $\downstate$ states from an incoherent mixture. 
The RF pulse at $t=0$ corresponds to rotation around the $y$ axis on a Bloch sphere and puts the BEC in a superposition state $| \theta \rangle = \sin(\theta/2)\upstate + \cos(\theta/2)\downstate$. Here $\theta = \Omega_R \Delta t$, where $\Omega_R$ is the RF Rabi frequency. In this state $P_z=-\cos(\theta)$ but the gas is still fully spin-polarised; the polarisation vector is $\vec{P} = \left ( \sin(\theta),0,-\cos(\theta) \right )$ and $P \equiv |\vec{P}| = 1$.
Subsequently the spin superposition decoheres, due to both intrinsic spin diffusion \cite{Widera:2008} and small magnetic field inhomogeneities \cite{phase_separation}.
$P_z$ is a constant of motion but the transverse polarisation decays and $P \rightarrow P_z$ [Fig.~\ref{fig:spin}(a)].

We study the transverse-polarisation decay in a Ramsey-type experiment.
Starting in the $\downstate$ state we apply two  $\pi/2$  RF pulses separated by time $t$ and then measure $P_z$. The first pulse creates a purely-transverse $\vec{P}=(1,0,0)$ and the second one maps the decaying $P$ into $P_z$ \cite{RFphase}.
As seen in Fig.~\ref{fig:spin}(b), we observe a very long spin-coherence time, $\tc \sim 10$~s. 
This means that in the unstable regime in Fig.~\ref{fig:surface} phase slips occur already at $t \lesssim \tc$, when we cannot equate $P$ and $P_z$.

We also perform a complementary experiment  
in which we adiabatically dress the rotating BEC with the RF field.
In presence of the RF field of frequency $\omega$, the effective magnetic field is $\vec{B}_{\rm eff} = (2 \hbar /\mu_B) (0, \Omega_R, -\delta)$, where $\mu_B$ is the Bohr magneton and  $\delta = \omega - \mu_B B/(2\hbar)$ is the detuning from resonance. On resonance, $\vec{B}_{\rm eff} \propto \hat{y}$. At $t=0$ we adiabatically (in 100 ms) sweep $\delta$ from a large value ($ \gg \Omega_R$) to zero, thus preparing a $P_z=0$ superposition state $|y\rangle = \left ( \upstate + i \downstate \right ) /\sqrt{2}$ [Fig.~\ref{fig:spin}(c)]. At this point, $|y\rangle$ is equivalent to the $|\pi/2\rangle$ state prepared by an RF pulse, which does not show long-term current stability [Fig.~\ref{fig:movie}(b)]. However, if we leave the RF field on during the in-trap evolution, $|y\rangle$ is an eigenstate of the Hamiltonian and the coherence between $\upstate$ and $\downstate$ components does not decay \cite{undressing}.
In this case $P_z=0$ supercurrent is stable and persists for more than a minute [Fig.~\ref{fig:spin}(d)].

These experiments clearly show that for analysing current stability in a partially polarised gas we must distinguish $P_z$ and $|\vec{P}|$.
With this understanding, we now quantitatively characterise the onset of the supercurrent decay in Fig.~\ref{fig:surface} by the time $\tau$ at which the probability that the first phase slip ($q=3 \rightarrow 2$) has occurred is $50\,\%$; this closely corresponds to the border of blue- and white-shaded regions.

\begin{figure*}[tbp]
\includegraphics[width=\textwidth]{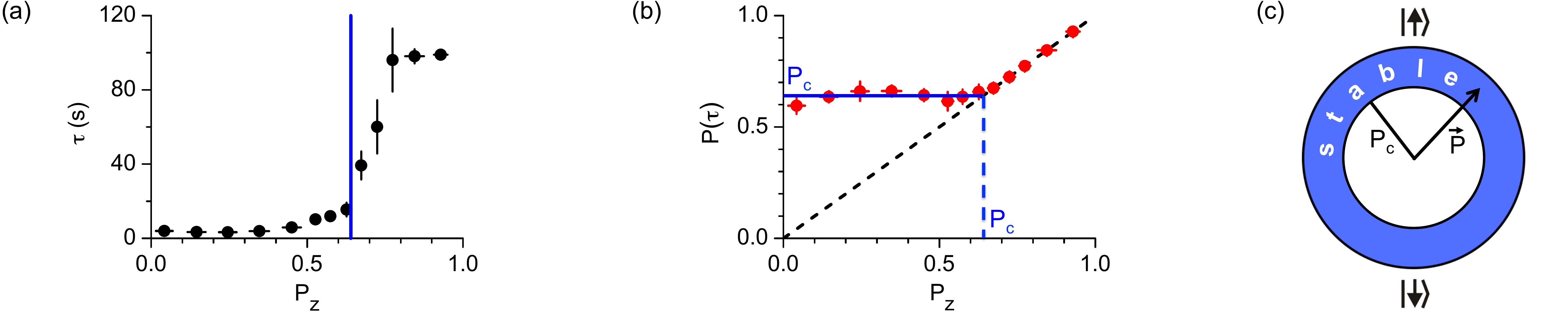}
\caption{(color online) Critical spin polarisation $P_c$.
(a) Characteristic time of the first phase slip, $\tau$, versus $P_z$. Vertical blue line marks $P_c$, accurately determined in (b).  
(b) $|\vec{P}|$ at the onset of the supercurrent decay. Fit to the points with $P(\tau)>P_z$ (horizontal blue line) gives $P_c = 0.64(1)$.
(c) Stability diagram on the Bloch sphere. The blue-shaded region in Fig.~\ref{fig:surface} maps into the outer shell $|\vec{P}|>P_c$.
}
\label{fig:Pc}
\end{figure*}

In Fig.~\ref{fig:Pc}(a) we see that $\tau$ rapidly increases for $P_z \gtrsim 0.64$, saturating at 100~s due to the $N$-decay.
We now combine our measurements of $\tau$ 
and the transverse-polarisation decay $f(t)$ [Fig.~\ref{fig:spin}(b)] to calculate $|\vec{P}|$ at the onset of supercurrent decay:
\begin{equation}
P(\tau) = \sqrt{ P_z^2 + (1-P_z^2) f(\tau) } \; .
\label{eq:P}
\end{equation}
In Fig.~\ref{fig:Pc}(b) we clearly distinguish two regimes: one where $P(\tau)$ is constant (within errors) and one where $P(\tau) = P_z$. We thus complete our physical picture and accurately determine the critical spin polarisation $P_c$:

(1) If $P_z > P_c$, then $|\vec{P}|$ {\it never} drops below $P_c$, the supercurrent is fundamentally stable, $\tau \gg \tc$ and $P(\tau) = P_z$.

(2) If $P_z < P_c$, supercurrent decay starts at  $\tau \lesssim \tc$, when the decaying $P$ becomes equal to $P_c$. 
From all the data in this regime we get $P_c = 0.64(1)$.

For $0\leq P_z \leq P_c$ the value of $\tau$ varies from $4\;$s to $15\;$s and the orientation of 
$\vec{P}(\tau) = ( \sqrt{P_c^2 - P_z^2}, 0, P_z )$ in spin space varies from purely transverse to purely longitudinal, but the onset of the supercurrent decay always occurs at the same $|\vec{P}|$.  
We thus conclude that the region of supercurrent stability is in fact the outer shell of the Bloch sphere where $|\vec{P}|>P_c$ [Fig.~\ref{fig:Pc}(c)]. This spin-rotational symmetry is intuitive but we note that it need not be universal. In our $^{87}$Rb gas the strengths of intra- and inter-component interactions are almost identical, so the Hamiltonian is almost invariant under rotations in spin space. In the future it would be very interesting to study supercurrent stability as a a function of both intra- and inter-component coupling strengths.

The existence of a critical population imbalance for superflow stability was predicted in Refs.~\cite{Smyrnakis:2009, Bargi:2010,Anoshkin:2012}, assuming equal intra- and inter-component interactions and no inter-component coherence. The current instability was associated with out-of-phase density fluctuations in the two components. 
However the agreement on the value of $P_c$ has not been reached. 
In Refs.~\cite{Smyrnakis:2009,Bargi:2010} it was predicted that any $q>1$ flow is unstable for essentially any $P<1$, but according to Ref.~\cite{Anoshkin:2012} such current is stable above some non-trivial interaction-dependent $P_c$. 
The latter conclusion qualitatively agrees with our observations.
However, none of the existing theories is quantitatively applicable to our experiments, since they are limited to the simplified cases of reduced dimensionality and very weak interactions.
Moreover, the interplay of the spin and rotational degrees of freedom may involve new physical effects. Specifically, the dynamics of the local spin vector on the Bloch sphere can result in a Berry phase and unwind the ``scalar" phase describing the rotational flow; the timescale for this process would be the same as the spin-decoherence seen in a Ramsey experiment \cite{Baur}.

In summary, we have observed persistent currents in multiply connected spinor condensates, demonstrated the existence of a critical spin polarisation for stable superflow, and elucidated the role of spin coherence in supercurrent stability. Our results should stimulate further theoretical work on this fascinating many-body problem and are also relevant for applications in trapped-atom interferometry.
An important next  step would be to study supercurrents in a two-species system with significantly different intra- and inter-component interactions.

\acknowledgments{We thank A. Gaunt and R. Smith for experimental assistance and N. Cooper, S. Baur, E. Demler, T. Kitagawa, G. Kavoulakis, J. Smyrnakis, J. Dalibard and J. Thywissen for discussions. This work was supported by EPSRC (Grants No. EP/G026823/1 and No. EP/I010580/1).}


\begin{thebibliography}{10}
\bibitem{symmetry}
Based on $\approx 500$ measurements we find essentially symmetric behaviour for negative $P_z$. 
One small difference is that no phase slips occur for about 70~s for $P_z \rightarrow -1$ and 90~s for $P_z \rightarrow 1$.
\bibitem{phase_separation}
In presence of in-plane field gradients the two-component gas is extremely susceptible to phase separation. By balancing $\upstate$ and $\downstate$ in-trap density distributions we minimised such gradients to $< 5 \times 10^{-5}$~G/cm.
\bibitem{RFphase}
Due to long separation the two RF pulses have a random relative phase $\varphi$. After the second pulse $P_z = P \cos (\varphi)$ and $\langle P_z^2 \rangle = P^2/2$. For each $t$ we  get $\langle P_z^2 \rangle$ from about 40 measurements.
\bibitem{undressing}
We checked for $t$ up to 100~s that we can always convert $|y\rangle$ into pure $\downstate$ by sweeping $\omega$ away from resonance.
\end{thebibliography}
\end{document}